\documentclass[12pt]{article}
\usepackage{amsmath}
\usepackage{amssymb}
\usepackage{epsf}

\newfont{\bit}{cmbxti10 scaled 1728}
\renewcommand{\thefootnote}{\fnsymbol{footnote}}

\begin{document}
\newpage
\pagestyle{empty}

\begin{center}
{\LARGE {Do Vortex Filaments in a Superfluid \\
Neutron Star\\
 Produce Gravimagnetic Forces ?\\
}}

\vspace{2cm}
{\large Herbert BALASIN
\footnote[1]{e-mail: hbalasin@tph.tuwien.ac.at}
\footnote[4]{supported by the APART-program of the 
Austrian Academy of Sciences}
}\\[.3cm]
{\em and}\\[.3cm]
{\large Werner ISRAEL
\footnote[2]{e-mail: israel@unix.uvic.ca}
\footnote[3]{supported by NSERC of Canada and by the Canadian
Institute for Advanced Research}
}\\[.5cm] 
{\em
Department of Physics and Astronomy,University of Victoria,\\
Victoria B.~C.~, V8W 3P6, CANADA
}\\[.8cm]
\end{center}
\vspace{1.5cm}
\begin{abstract}
A general analysis of the gravitational dynamics of a medium with
a continuous distribution of vorticity indicates that the answer to the 
question raised in the title is affirmative, contrary to a recent claim.
\end{abstract}
\vspace{1.3cm}

\rightline{TUW 97 -- 04}
\rightline{January 1997}
\vfill


\renewcommand{\thefootnote}{\arabic{footnote}}
\setcounter{footnote}{0}
\newpage  
\pagebreak
\pagenumbering{arabic}
\pagestyle{plain}
\renewcommand{\baselinestretch}{1}
\small\normalsize 
\section*{\large\bit 1) Introduction}

Very soon after the formation of a neutron star,
thermal energies inside it drop below the energy gap associated with 
neutron pairing. One therefore expects all but the outermost parts of 
the crust to pass rapidly into a superfluid state \cite{Mi,Gi,PiAl}.
The critical angular velocity for vortex formation is negligibly small for
bodies of astronomical size and therefore, as first noted by Ginzburg and 
Kirzhnits \cite{GiKi}, the star's interior will be threaded by vortex lines
parallel to the axis of rotation.\par
A stimulating review by Kirzhnits and Yudin \cite{KiYu1} discusses some of 
the interesting and surprising general-relativistic effects associated with
superfluid neutron stars. Consider, for instance, an imaginary experiment
in which the star is initially at rest and a spin is imparted to the crust. 
Lense-Thirring (frame-dragging) effects will make it appear that the 
superfluid interior has been set into partial co-rotation, i.~e.~, the 
contravariant azimuthal velocity $v_s^\phi = d\phi /dt$ with
respect to distant stationary observers will become non-zero.
If the crust is separated from the fluid by a small gap, then this induced 
rotation can easily exceed the critical angular velocity without forming any 
vortices. This is because the superfluid's orbital angular momentum, 
proportional to the covariant component $v_{s\phi}$, remains zero.
The induced rotation also makes no contribution to the value of the 
metric coefficient $g_{\phi t}$ (the ``gravimagnetic potential''),
whose non-vanishing is due entirely to the rotation of the crust.\par
Superfluid flow is in all circumstances irrotational.
For regular axisymmetric flows this implies $v_{s\phi}=0$, 
i.~e.~, the superfluid can never acquire orbital angular momentum. 
It is nevertheless able to entrap
angular momentum within the cores of its vortex filaments, which consist of
normal fluid. To simplify the description, one generally assumes that
it is permissible to replace the vortex array by a fictitious 
``effective'' orbital flow ($v^{eff}_{s \phi}\neq 0$) when averaging 
over macroscopic domains, i.~e.~, domains large compared with the vortex 
spacing (about $10^{-2}cm$ for the Crab pulsar). The equivalence
assumption is that gravitational and other physical effects of the vortex 
array are macroscopically indistinguishable from an effective orbital motion 
having the same distribution of angular momentum. This seemingly innocent 
assumption is called into question  in a recent analysis by Kirzhnits and
Yudin \cite{KiYu2}. The authors do not succeed in finding an acceptable
solution representing the contribution of a vortex line to the external
gravitational field of a spherical body in the linearized Einstein theory.
They conclude: ``The general formula relating the angular momentum
of a body with the asymptotic behavior of its gravimagnetic field
is invalid when vortex filaments are present.''\par
This conclusion is not easy to understand intuitively, since the 
formation and slow migration of vortices into the superfluid from the 
interface with the crust is macroscopically an axisymmetric process, which
conserves angular momentum locally and involves no appreciable
redistribution of mass. It is therefore hard to see how it can affect
the external gravitational field at all, much less suppress all
gravimagnetic effects, as the angular momentum of the crust gradually becomes
``vorticized''. In short, there is no obvious reason why spin and
orbital angular momentum should behave in a fundamentally different way
as sources of gravity.\par
The issue is perhaps elementary but clearly important because of the
potential implictaions for the gravitational dynamics of the binary pulsar 
and gravitational radiation from neutron star mergers. 
To remove any possible doubt it seems worth devoting some effort
to an explicit verification that bulk rotation and vortex filaments
are indeed equivalent in their gravitational effects, and this
is the object of our paper. In section 2 we review the phenomenological
form taken by the Einstein field equations for a medium which is 
gravitationally polarized by alignment of internal spins. The effect of this 
spin polarization on the gravimagnetic potential near infinity can be inferred
from an integral identity due to Komar, and this is discussed in section
3. These results are exact and valid for strong fields. If the gravitational
field is weak (i.~e.~non-relativistic) the linearized Einstein equations 
provide a complete solution of the problem (section 4). Finally section
5 is devoted to the simplest realisation of the superfluid 
neutron star within the framework of linearized gravity, namely a
spherical star of constant density with purely azimuthal, locally
irrotational flow. This flow corresponds to having a single vortex
filament along the axis. The origin of the contribution to the far-field 
gravimagnetic field is traced back to the singularity of the flow along 
the axis of rotation.

\section*{\large\bit 2) Einstein field equations for a medium with 
internal spin}

Analogously to the Maxwell-Lorentz equations for a dielectric medium,
there is a statistically averaged form of the Einstein field equations 
appropriate for the coarse-grained description of a gravitationally
polarized granular medium \cite{Isr1}. In the case of interest to us here
the polarization is a spin alignment of the granules.\par
At the sub-macrolevel, the material inside each granule is described by a
symmetric conserved tensor $T^{ab}_{micro}$, representing the
flux of 4-momentum, which couples to the Einstein tensor in the
standard way. On larger scales, it becomes meaningful to split the 
complex motion of the material into a translational motion of the granules and
a distribution of internal spins. Correspondingly the coarse-grained average
$T^{ab}= <T^{ab}_{micro}>$ splits into $\mathcal{T}^{ab}$, the flux of
translational 4-momentum and a contribution from $S^{abc}$, the flux
of spin angular momentum (skew in the first two indices: 
$S^{abc}=S^{[ab]c}$). Detailed analysis \cite{BaiIsr} shows
that the coarse-grained form of the Einstein field equations is
\begin{equation}\label{macein}
  G^{ab}=8\pi T^{ab}\quad\text{where}\quad T^{ab}={\mathcal T}^{ab} +
  \nabla_c U^{abc},\quad U^{abc} = \frac{1}{2}(S^{acb} + S^{bca} - S^{abc})
\end{equation}
The translational momentum flux ${\mathcal T}^{ab}$ is neither 
symmetric nor in general (in a curved spacetime) conserved. The symmetry and
conservation  of $T^{ab}$ are equivalent to the conditions
\begin{equation}\label{sym}
  {\mathcal T}^{[ab]} = \frac{1}{2}\nabla_c S^{abc},\quad 
  \nabla_b{\mathcal T}^{ab} = -\frac{1}{2} R^a{}_{bcd}S^{cdb}.
\end{equation}
Equations (\ref{macein}) are formally identical with the well-known 
Belinfante-Rosen\-feld prescription for symmetrizing what is called 
(in a field-theoretic context) the ``canonical'' stress-energy tensor 
\cite{Wein1}.\par
The simplest illustrative example is a gas of spinning particles. If
the skew tensor $s^{ab}$ denotes the angular momentum of a particle,
$p^a$ its 4-momentum and $v^a$ its 4-velocity, then
$$
{\mathcal T}^{ab}(x) = \int N(x,p,s) p^a v^b \omega_g(p,s), 
\qquad S^{abc}(x) = \int N(x,p,s) s^{ab}v^c \omega_g(p,s),
$$
where $N=N(x,p,s)$ is a distribution function and $\omega_g(p,s)$ the 
invariant volume form on spin-momentum space. Here the asymmetry of 
${\mathcal T}^{ab}$ arises from the non-collinearity of $p^a$ and $v^a$
for a spinning particle.

\section*{\large\bit 3) Tolman-Komar integral identities}

To check that the spin angular momentum concealed in (\ref{macein})
makes its proper contribution to the gravimagnetic potential $g_{\phi t}$,
it is simplest to appeal to the Tolman-Komar integral identities
(see e.~g.~\cite{FroIsrUnr}. These are rigorous general-relativistic 
analogues of Gauss's flux theorem, which hold for spacetimes admitting
Killing-symmetries and relate volume integrals over the matter content
of an isolated system to surface integrals at infinity
\begin{align}
&\nabla_{(a}\xi_{b)} =0 \quad\Rightarrow\quad 
\nabla^2\xi_b = - \xi_a R^a{}_b\nonumber\\
&\int\limits_{\partial\Sigma}\nabla_a\xi_b d^2\sigma^{ab} =
-8\pi \int\limits_{\Sigma} \xi_a(T^a{}_b -\frac{1}{2} 
\delta^a{}_b T) d^3\sigma^b\nonumber\\
\end{align}
For stationary fields ($\xi^a=\partial_t^a$) we have Tolman's theorem
$$
m = \int(T^i{}_i -T^t{}_t)\sqrt{-g}d^3x,
$$
where the integration is over a $t=const$ 3-space, $i$ runs over the three 
spatial coordinates, and the gravitational mass $m$ is defined by the 
asymptotic form  
$$
g_{tt}\approx -(1-\frac{2m}{r}) \qquad (r\to\infty).
$$
For an axisymmetric system, the gravitating angular momentum $ma$,
defined by the asymptotic form 
$$
g_{\phi t}\approx -\frac{2ma\sin^2\theta}{r} \qquad (r\to\infty)
$$
in asymptotically spherical coordinates, is similarly given by
\begin{equation}\label{spin}
ma = \int T_\phi{}^t \sqrt{-g}d^3x,
\end{equation}
For our purpose  $T_\phi{}^t$ is most conveniently recast
in the covariant form $\xi^a T_a{}^b \nabla_b t$, 
where $\xi^a=\partial_\phi^a$ is the axial Killing vector. 
Making use of (\ref{macein}), the identity
\begin{equation}\label{spinid}
  U^{abc}-U^{cba}=S^{abc},
\end{equation}
and the Killing property that $\nabla_a\xi_b$ is skew, we find
$$
\int\limits_\Sigma \xi^a(T_a{}^b - {\mathcal T}_a{}^b)d^3\sigma_b =
\int\limits_\Sigma \nabla_c(\xi^a U_a{}^{bc}) d^3\sigma_b - 
\frac{1}{2}\int\limits_\Sigma \nabla_c\xi_a S^{acb} d^3\sigma_b
$$
Since the total derivative integrates to zero, (\ref{spin}) becomes
\begin{equation}\label{spinorb}
  ma = \int{\mathcal T}_\phi{}^t\sqrt{-g}d^3x - \frac{1}{2}
\int S^{act}(\nabla_c\xi_a) \sqrt{-g}d^3x,
\end{equation}
in which the separate contributions of the orbital and
spin angular momentum are displayed explicitly.
To check that the second term has indeed the right form, we go to the 
non-relativistic weak-field limit, in which we can set
$$
\xi_a \approx \rho^2\nabla_a\phi\qquad S^{\rho\phi t}\approx\frac{1}{\rho}
\epsilon^{\rho\phi z} S_z,
$$
in cylindrical coordinates ($\epsilon^{abc}$ is the three-dimensional 
permutation symbol).
Then the second term in (\ref{spinorb}) reduces to $\int S_zd^3x$,
giving the $z$-component of the total spin angular momentum as it should.

\section*{\large\bit 4) Linearized theory}

In the case of weak fields we can obtain a complete solution to the
problem by integrating the linearized Einstein equations \cite{Wein2}.\par
In approximately rectangular coordinates the metric perturbation $h_{ab}$
and its trace-reversed form $\bar{h}_{ab}$ are ``small'' quantities
defined by
$$
g_{ab}=\eta_{ab} + h_{ab}\qquad \bar{h}_{ab} = h_{ab} - \frac{1}{2}h
\eta_{ab},
$$
where $\eta_{ab}$ is the Minkowski metric. The coordinate
freedom  allows us to impose the four (``harmonic'') gauge
conditions $\partial_a\bar{h}^{ab}=0$, with respect to which the Einstein
equations and their integrability condition (the linearized 
conservation laws) reduce to the simple form
$$
G_{ab} = -\frac{1}{2}\partial^2\bar{h}_{ab} = 8\pi T_{ab}, \qquad
\partial_aT^{ab}=0,
$$
where $\partial^2$ denotes the Minkowski wave operator. We specialize to 
stationary fields where the solution is
\begin{equation}\label{sol}
 \bar{h}_{ab}(x) = 4 \int T_{ab}(x')\frac{1}{|x-x'|} d^3x',
\end{equation}  
and the conservation laws imply that (to linear order) the sources must obey 
the constraints
\begin{equation}\label{lincon}
  \int J_i d^3x = \int T_{ij} d^3x =0, \qquad \int(x^iJ^k +x^kJ^i)d^3x =0
\end{equation}
where $J^i = T^{it}$ is the momentum density and the indices $i,k$ run
from 1 to 3. 
(For instance the last of these follows from  the integration of 
$\partial_i(x^i x^k J^l)$ over a bounded source and using 
$\partial\cdot J =0$.)\par
The external field of a compact source, at distances large compared to its 
dimensions, is thus given by
$$
\bar{h}_{ij}=0,\quad\bar{h}_{tt}= \frac{4m}{r}.
$$
To obtain the gravimagnetic vector-potential $A_i =\bar{h}_{it}= h_{it}$
we expand $|x-x'|^{-1}$ in (\ref{sol}) up to the first order in $x'$
and note (\ref{lincon}). This yields $A_i$ in terms of the source's angular
momentum $L_i$:
\begin{equation}\label{gravpot}
  A_i (x) = 2r^{-3} (x \times L)_i,\qquad L_i = \int (x\times J(x))_i d^3x.
\end{equation}
For an axisymmetric source it is convenient to switch to spherical
polar coordinates, for which 
$$
A_i dx^i = -2r^{-3} (L^i x^k\epsilon_{ikl}dx^l) = 
-\frac{2ma}{r}\sin^2\theta d\phi,
$$
where we introduced the conventional notation $L_z=ma$. Putting all this 
together, we arrive at the standard form of the linearized stationary 
exterior metric
$$
ds^2 = (1+\frac{2m}{r})(dx^2 + dy^2 + dz^2) -\frac{4ma}{r}\sin^2\theta
d\phi dt - (1-\frac{2m}{r})dt^2.
$$
To exhibit directly the separate contributions of spin and orbital
angular momentum to the gravimagnetic potential in (\ref{gravpot}),
let us assume the orbital 4-velocity $u^a$ non-relativistic,
so that we can set $u^t\approx 1$ and neglect $u^iu^i$.
To this order we may set
$$
S^{abc}=S^{ab}u^c,\qquad S^{ab}u_b=0,
$$
which implies $S^{it}=S^{ik}u_k$. From (\ref{macein}) we then find to 
first order in $u^i$
$$
2U^{itk} = S^{ik} = \epsilon_{ikl}S_l,\qquad \partial_kU^{itk}=\frac{1}{2}
(\partial \times S)^i.
$$
The total momentum current $T^{it}$ in (\ref{macein})
therefore decomposes as
$$
J^i = \rho u^i + \frac{1}{2}(\partial \times S)^i.
$$
Substituting in (\ref{gravpot}) finally yields the expected result
$$
L^i = \int (x \times \rho u)^i d^3x + \int S^i d^3x,
$$
verifying that the angular momentum in vortices contributes in the 
same way as orbital angular momentum to the gravimagnetic potential.

\section*{\large\bit 5) A simple example: gravimagnetic field of a single
vortex filament on the axis of a homogenous star}

In order to give a concrete realization of the situation discussed
in the previous sections let us consider the simplest possible model by
taking the superfluid velocity to be the irrotational 
azimuthal flow $d\phi$ and assuming the mass-density to be constant 
within the star.
The equation for the gravimagnetic potential (within the linearized
theory) become
\begin{equation}\label{gravimag}
*d{*d}A = *d{*F} = j = \theta(a-r) d\phi
\end{equation}
The fact that the ``velocity'' $d\phi$
becomes singular at the axis may be conveniently described by 
taking the distributional identity
\begin{equation}\label{domega}
d\phi = \frac{xdy-ydx}{x^2+y^2} =: \omega_\phi \qquad
d\omega_\phi = 2\pi \delta^{(2)}(x) d^2x 
\end{equation}
into account, where $\omega_\phi$ denotes the tensor-distribution on 
${\rm I\!R}^3$, corresponding to $d\phi$. It is not closed and therefore 
certainly not exact. This fact will actually be important in the description
of a vortex filament along the axis of rotation.
Integration of the equation for the gravimagnetic potential 
(\ref{gravimag}) yields
\begin{align}\label{field}
 &d{*F}= *j = \theta(a-r)dr\frac{d\theta}{\sin\theta}\nonumber\\
 &{*F}= \omega_0 + \theta(a-r)(r-a)\frac{d\theta}{\sin\theta}
\qquad d\omega_0 =0 \Rightarrow \omega_0=df\nonumber\\
 &F= *df - \theta(a-r)(r-a) drd\phi 
\end{align}
where $\omega_0$ denotes an ``integration constant'' (solution 
of the homogeneous equation).  Since the second term in (\ref{field})
naively appears to be closed, one might be tempted to discard $\omega_0$.
However, in the light of (\ref{domega}) we find 
\begin{align}\label{fil}
&dF=0=d{*d}f + \theta(a-r)(r-a)dr \> 2\pi\delta^{(2)}(x) d^2x,\nonumber\\
&\Delta f = *d{*df} = -2\pi\theta(a-|z|)(|z|-a)\frac{z}{|z|}\delta^{(2)}(x).
\end{align}
This shows that we cannot take $\omega_0$ to be zero,
which would have produced a gravimagentic field of compact support,
namely within the star.
It is precisely the existence of the filament, represented by concentrated
terms in (\ref{fil}), which gives rise to contributions which are not of 
compact support
and therefore maybe detected in the far-field. Since our model is
sufficiently idealized it is possible to solve the Laplace-equation
for $f$ exactly.  The calculation is considerably facilitated by
taking a $z$-derivative of $\Delta f$, which simplifies
the inhomogeneity
$$
\Delta \partial_z f = 4\pi a\delta^{(3)}(x) - 2\pi\theta(a-|z|)
\delta^{(2)}(x),
$$
and gives rise to the solution
\begin{align}\label{potential}
  f &= -a \log\left(\frac{z}{\rho} + \sqrt{\frac{z^2}{\rho^2}+1} \right)
- \frac{1}{2}\left[ 
 (z-a)\log\left(\frac{z-a}{\rho} + \sqrt{\frac{(z-a)^2}{\rho^2}+1} \right)
\right.\nonumber\\
 &- \sqrt{(z-a)^2 + \rho^2} 
- (z+a)\log\left(\frac{z+a}{\rho} + \sqrt{\frac{(z+a)^2}{\rho^2}+1} \right)
\nonumber\\
 &\left. + \sqrt{(z+a)^2 + \rho^2}
\right].
\end{align}
Inserting $\omega_0=df$ into (\ref{field}) turns the expression for $F$ 
into
\begin{align}\label{gravmagfield}
F &= \frac{-a}{\sqrt{\rho^2+z^2}}\left( dz-\frac{z}{\rho^2}\tilde{x}
d\tilde{x} \right) -\frac{1}{2}\left( dz 
\log\frac{z-a + \sqrt{(z-a)^2+\rho^2}}{z+a + \sqrt{(z+a)^2+\rho^2}}\right. 
\nonumber\\
&\left. + \left(\sqrt{(z+a)^2 + \rho^2}-\sqrt{(z-a)^2+\rho^2)}\right)
\frac{\tilde{x}d\tilde{x}}{\rho^2}\right) - \theta(a-r)(r-a) drd\phi.
\end{align}
Integration of (\ref{gravmagfield}) gives rise to the potential $A$

\begin{align}
A &= \left\{ \frac{\rho^2}{4} \log\frac{z+a + \sqrt{(z+a)^2+\rho^2}}
{z-a + \sqrt{(z-a)^2+\rho^2}} + \frac{z+a}{4} \sqrt{(z+a)^2+\rho^2} 
\right.\nonumber\\
&\left. \hspace*{1cm}-\frac{z-a}{4} \sqrt{(z-a)^2+\rho^2} -a r -\frac{1}{2}
\theta(a-r)(a-r)^2  \right \} d\phi
\end{align}
which is not zero in the asymptotic region.

\section*{\large\bit Conclusion}
In this paper we reviewed the influence of spin and orbital angular
momentum of a material system on its gravitational field. Within
the general framework of gravitationally ``polarized'' matter
one can show that both of them contribute to the angular momentum
detected by an asymptotic observer. This investigation settles the 
recently discussed issue about the imprint of vortex lines found in 
a superfluid neutron star on the asymptotic gravitational field. 
Specifically in the linearized approximation of general relativity, 
one can obtain an explicit solution which clearly exhibits the
effect of the vortex line on the gravimagnetic potential. 
\vfill
\noindent
{\bf Acknowledgement:} We would like to thank George Vlasov for
stimulating correspondence.

\end{document}